\documentclass{report}
\usepackage{ohm-ngc}
\usepackage{multirow}
\usepackage{booktabs}
\usepackage{color}
\usepackage{graphicx}
\usepackage{amsfonts}
\usepackage{amsmath}
\usepackage{subcaption}
\usepackage{fancyhdr}

\begin{document}

\title{Seed Selection for Spread of Influence in Social Networks: Temporal vs. Static Approach}

\author{Rados{\l}aw MICHALSKI, Tomasz KAJDANOWICZ, Piotr BR\'{O}DKA, Przemys{\l}aw KAZIENKO}{Institute of Informatics, Wroc{\l}aw University of Technology, Wybrze{\.z}e Wyspia{\'n}skiego 27, 50-370 Wroc{\l}aw, Poland}

\E-mail{\{radoslaw.michalski, tomasz.kajdanowicz,\\piotr.brodka, kazienko\}@pwr.edu.pl}
\thispagestyle{empty}

\fancypagestyle{empty}{%
  \fancyhf{}
  \fancyfoot[C]{Preprint, this article has been published in\\New Generation Computing, Vol. 32, Issue 3-4, pp. 213-235\\Ohmsha-Japan and Springer (2014), DOI 10.1007/s00354-014-0402-9}
}

\begin{abstract}
The problem of finding optimal set of users for influencing others in the social network has been widely studied. Because it is NP-hard, some heuristics were proposed to find sub-optimal solutions. Still, one of the commonly used assumption is the one that seeds are chosen on the static network, not the dynamic one. This static approach is in fact far from the real-world networks, where new nodes may appear and old ones dynamically disappear in course of time.

The main purpose of this paper is to analyse how the results of one of the typical models for spread of influence - linear threshold - differ depending on the strategy of building the social network used later for choosing seeds. To show the impact of network creation strategy on the final number of influenced nodes - outcome of spread of influence, the results for three approaches were studied: one static and two temporal with different granularities, i.e. various number of time windows. Social networks for each time window encapsulated dynamic changes in the network structure. Calculation of various node structural measures like degree or betweenness respected these changes by means of forgetting mechanism - more recent data had greater influence on node measure values. These measures were, in turn, used for node ranking and their selection for seeding. 

All concepts were applied to experimental verification on five real datasets. The results revealed that temporal approach is always better than static and the higher granularity in the temporal social network while seeding, the more finally influenced nodes. Additionally, outdegree measure with exponential forgetting typically outperformed other time-dependent structural measures, if used for seed candidate ranking. 
\end{abstract}


\section{Introduction}
While studying the evolution of social network analysis research\cite{prell2011social}, it is clearly observed that at the beginning researchers focused on analysing multiple static\cite{kazienko2012} or aggregated networks. They have provided very challenging set of research problems closely related to graph theory: computation of betweenness centrality measure\cite{Freeman1977} based on shortest paths in the network\cite{spira1975finding} or group discovery methods\cite{PallaBarabasiVicsek2007} by using clique finding algorithms\cite{moon1965cliques}. Due to computational complexity limitations, most of the dynamic processes analysed in the networks were modelled on basic, static network structures. An alternative approach, recently extensively explored, are temporal networks, i.e. networks that reflect the occurrence of events in time\cite{holme2012temporal} and changes in their sets of nodes and edges. Moreover, the network structural dynamics may also be reflected in other dynamic processes, such as information diffusion or spread of influence. These processes can be strongly influenced by appearing and disappearing nodes and their dynamic relationships and this complex phenomena attracted more and more researchers worldwide. A simple proof for that is the number of full text articles containing the term "temporal social networks" in scientific databases, e.g. in Scopus\footnote{Elsevier's Scopus Database - http://www.scopus.com}, this number has been increasing on average by about 35\% every year since 2005.

The approach to network dynamics with the highest granularity is a separate consideration of each individual event log resulting in changes in the network structure. In such case, each edge in the network must be timestamped and may overlap with many other edges for other time points. This idea, however, is the most computational and storage demanding, so to overcome its limitations, two main network aggregation types may be applied: static and temporal. The former is very popular and quite simple: all recorded events are aggregated into a single static network, e.g. if two users exchanged at least one email any time over last two years, they are treated as mutually connected acquaintances. In this paper, such concept will be further called \textit{a static approach} since it losses the temporal nature of timestamped events. Some recent research have revealed that this may result in misleading conclusions about the outcomes of dynamic processes\cite{MasudaHolme2013}. Moreover, it has been shown that the social spreading phenomenon is very dependent on the timing of contacts\cite{karsai2011small}. As a result of these findings, the research can turn towards temporal networks that only partially aggregate events and provide a number of time-ordered networks; each merges events for a particular period - time frame. This approach, which may be considered as a trade-off between working with the static network and event log. It facilitates benefiting from all the achievements of graph theory but it simultaneously results with less storage requirements and computational complexity than raw event log processing.

The main purpose of this paper is to analyse how the results of one of the typical models of spread of influence - linear threshold\cite{Granovetter1978} - differ depending on the strategy of building the social network used later for choosing seeds. To show the impact of network building strategy on spread of influence results three kinds of approaches were studied: one static and two temporal. These concepts were also utilized in experimental verification on five empirical datasets. For all of the networks, the authors used the most popular heuristics for choosing initial seeds for spread of influence based on network measures and they observed the results of the process. Additionally, some new heuristics were proposed, which consider the role of time in evaluating nodes as potential seeds.


\section{Related Work}
Apart from empirical studies on the spread of influence\cite{rogers2010diffusion}, a number of theoretical models of this process were proposed and widely studied. These include among others: linear threshold model LT\cite{Granovetter1978}, independent cascade model IC\cite{goldenberg2001talk} or the voter model VM\cite{clifford1973votermodel}. Each of them introduces different approaches of modelling the process. For instance, linear threshold model is oriented on the percentage of influenced neighbours of a node, while independent cascade model introduces the probabilities of influencing nodes assigned to node which is already influenced. These principles of social propagation mechanisms were presented in\cite{krolprop}.

One of the most interesting research questions is the problem of maximising the final spread of influence as defined by Kempe et al.\cite{KempeKleinbergTardos2003}. They considered the question, which nodes should be chosen for a particular spreading model to maximize the overall number of influenced nodes. Due to the fact that the problem is NP-hard for most models, so far, only some heuristics were proposed based on different approaches and various initial assumptions. In\cite{mathioudakis2011sparsification} the authors studied how to select initial nodes in an independent cascade model based on the static graph and the previous propagation logs, which were assumed to be known a priori. Some other scalable algorithms for finding seeds for the LT model were suggested in\cite{chen2010scalable} and\cite{GoyalLuLakshmanan2011}, while maximising the spread for the IC model was considered in\cite{SaitoNakanoKimuraLovrekHowlettJain2008}. Since most of the studies were devoted to LT and IC models, it is also worth to mention about some research on maximising the spread of influence in the voter model\cite{Even-DarShapira2007}. Meanwhile, an interesting approach was presented in\cite{GoyalBonchiLakshmananVenkatasubramanian2012}, where other versions of the problem were analysed: minimising the time of the spread of influence or minimising the budget for finding seeds. It is worth to mention that some researchers also try to overcome the limitations of basing just on the graph structure to find influential nodes. However again, most of these approaches operate on a static view of the social network, which is the strong simplification of the reality, even when considering multiple layers\cite{convince}. In\cite{goyal2011data} authors use historical data to calculate the propagations probabilities, however the limitation of this method is that this historical data should be known in advance. Masuda and Holme stated, while studying epidemic processes\cite{MasudaHolme2013}, that the static approach of the network epidemiology may miss a great deal of what happens in the long-term reality, since the nodes contact each other only in particular time windows. Similarly, Gomez-Rodriguez et al.\cite{Gomez-Rodriguez_Leskovec_Krause_2010} confirmed that relaxing the assumption of static propagation network would be an interesting case for further research in the area of spread of influence. Kossinets and Watts also emphasized the importance of time in analysing the processes in social networks\cite{Gueorgi2006}.

Having these conclusions in mind, it was decided to evaluate how the most typical heuristics based on the network structure perform on temporal and static networks. A special attention has been turned to observation of dynamics of the influence that spreads over temporal (changing) network after choosing initial seed sets. Indeed, this direction is emerging, because first work studying spreading phenomenon in temporal networks is published nowadays\cite{jankowski2013,karimi2013threshold}. However, the authors did not focus on seeding strategies there, but they analysed the process under different assumptions for the aggregation level.

\newpage
\section{General Concept}

\subsection{Linear Threshold Model}

Let us consider spread of influence within the framework of temporal social network. A temporal social network (TSN) consists of interval graphs $TSN=<G_1, G_2, \ldots, G_l, \ldots, G_K>$, where graph nodes and edges correspond to nodes' social common activity in a given interval out of the set of intervals  $T_K = \{(t_1, t_1'), \ldots, (t_l, t_l'), \ldots, (t_K, t_K')\}$, $K \in \mathbb{N}_+$ called time windows. The parentheses $(t_l, t_l')$ indicate the period of activity, the unprimed time marks the beginning of the window and the primed quantity denotes its end\cite{holme2012temporal}. There exists an edge $(v_i, v_j)$ in a particular interval graph $G_l$, $1\leq l \leq K$ if and only if there exists a social relationship between $v_i$ and $v_j$ in $(t_l, t_l')$ time window. Due to the fact that there might appear multiple events (common social activities) between $v_i$ and $v_j$ within a single time window $l$, the relationship is obtained by aggregation of these events. Therefore, each interval graph $G_l$ can be treated as a static graph.

Each interval graph $G_l$, $1\leq l \leq K$, is composed of a set of nodes $V_l=\{v_1, v_2, \cdots, v_n\}$ and a set of directed edges $E_l$ representing relations between nodes in time window $l$: $E_l=\{(v_i,v_j)|v_i,v_j\in V_l\}$. Let $N_i(G_l)$ be the set of directly neighbouring and potentially influencing nodes, i.e. nodes with relation to node $v_i \in V_l$ in window $l$: $N_i(G_l)=\{v_j|(v_j,v_i) \in E_l\}$. In other words, the set $N_i(G_l)$ is composed of individuals who can potentially influence node $v_i$ in time window $l$.

It is assumed that before the spread of influence (before time window $l=1$), a subset of individuals $\Phi(0) \subset V_0$ is selected as the seed for the influence spread. By $V_0$, we denote a set of all nodes that had been observed in the network before an influence spread was considered.

The set $\Phi(0)$ should represent a group of individuals who have already been influenced as well as the set of promoters who have certain social, economic and/or political abilities to influence others. 

\newpage
We assume that the initial seed set adopts all influence before observed spread starts. At the following time window $(t_1, t_1')$, an individual $v_i \in V_1 \setminus \Phi(0)$ will be really influenced if at least $\phi_i \in (0,1]$ fraction of its neighbours are in the seed set, i.e.

\begin{equation}
\frac{\left \vert{\Phi(0)\cap N_i(G_1)}\right \vert}{\left \vert{N_i(G_1)}\right \vert}\geq\phi_i \Rightarrow v_i \in \Phi(1)
\end{equation}

It means that set $\Phi(1)$ consists of all nodes who have been exposed to the influence, are persuaded by their neighbours and adopted the influence in period $(t_1, t_1')$.

In general, for a given $k \in \mathbb{N}$, a not-yet-influenced node $v_i \in V_k \setminus \bigcup_{l=0}^{k-1}\Phi(l)$ will be influenced in the $k$th window $(t_k, t_k')$, if

\begin{equation}
\label{Eq:kthInfluenceStep}
\frac{\left \vert{\{\bigcup_{l=0}^{k-1}\Phi(l)\}\cap N_i(G_k)}\right \vert}{\left \vert{N_i(G_k)}\right \vert}\geq\phi_i \Rightarrow v_i \in \Phi(k)
\end{equation}

Finally, we obtain a list of nodes influenced in the following periods: $\Phi(0), \Phi(1), \dots, \Phi(l), \dots, \Phi(K)$.

According to Eq. (\ref{Eq:kthInfluenceStep}), the final set of influenced nodes $\bigcup_{l=0}^{K}\Phi(l)$ depends on two crucial factors: initial seed set $\Phi(0)$ and the dynamics of consecutive interval graphs $G_l$ for $1\leq l\leq K$ stating the influencing neighbourhoods $N_i(G_l)$ for each node $v_i$ in the consecutive periods. 

\subsection{Spread of Influence in the Temporal Social Network}

Regardless the propagation model used for spread of influence, the seed selection strategy determines the final number of influenced nodes in the network. Given the temporal social network $TSN$ that consists of $K$ interval graphs, the goal is to select initial seed set of nodes $\Phi(0)$ of size $m$ $(|\Phi(0)| = m)$ in order to maximize the final number of influenced nodes after the $K$th window $\Phi(K)$ in the influence propagation process, see Eq. \ref{Eq:argmax}. 

\begin{equation}
\label{Eq:argmax}
\arg\max_{\Phi(0),|\Phi(0)| = m} 
\left \vert{\Phi(0) \cup \bigcup_{l=1}^{K} \left\{{v_i:v_i \in V_l \setminus \bigcup_{k=0}^{l-1}\Phi(k) \wedge \frac{|{\bigcup_{k=0}^{l-1} \Phi(k)}\cap N_i(G_l)|}{|N_i(G_l)|}\geq \phi_i}\right\}}\right \vert
\end{equation}
Due to the fact that the final set of influenced nodes $\bigcup_{l=0}^{K}\Phi(l)$ depends on initial seed set $\Phi(0)$ and the interval graphs $G_l$ for $1\leq l\leq K$, its exact estimation is highly complex. Under the LT propagation model, it has been already shown to be NP-hard\cite{KempeKleinbergTardos2003}. 

The real setting of influence spread problem might be much more complicated. In real world applications, it can be expected to select seeds using historical knowledge about nodes activity. In this case the information about past activity of nodes may become an advantage, because, for instance, recently inactive nodes may be omitted from the seeds set improving general results.

Still, the dynamics of the complex networks introduces completely new problems in comparison to the static approach. In particular, we would need to address the following problems:
\begin{itemize}
	\item As the activity of nodes may differ, even highly active nodes may become inactive just after the moment of seed selection. If a node is chosen as a seed, it is expected that it will be active also later on, which may be not necessarily true leading to wasting the marketing campaign budget.
	\item After the initial influencing moment, the increasing dynamics of the network in terms of appearing new nodes may minimize the expected influence of old nodes chosen as seeds.
	\item Due to the fact that the network dynamics also includes changes in edges, it may happen that kind of dynamics may be either helpful or harmful, i.e. influential nodes meet not previously expected non-influenced nodes, but it may also lead to undesirable outcomes - the expected node behaviour (high susceptibility to influences or high ability to influence others) may not necessarily be valid any more.
\end{itemize}

In fact, the above mentioned problems could be solved, if the link prediction solutions\cite{liben} would foresee new links with acceptable level of accuracy. This, in turn, could enable development of completely new seeding methods for dynamic networks, yet, still there is a lot of research to be performed before it comes true.

\subsection{Research Problem: How to Select the Initial Seed?}
The main problem investigated in this paper is how to select initial set of nodes that next could influence others in the most efficient way. The efficiency is here measured by the number of finally influenced nodes. The main assumption introduced here is that the social network, in which the influence spreads is dynamic, i.e. nodes and social relationships may both appear and disappear. This dynamic phenomena of the network may significantly influence the final outcome. 

\begin{figure}
\includegraphics[width=\columnwidth]{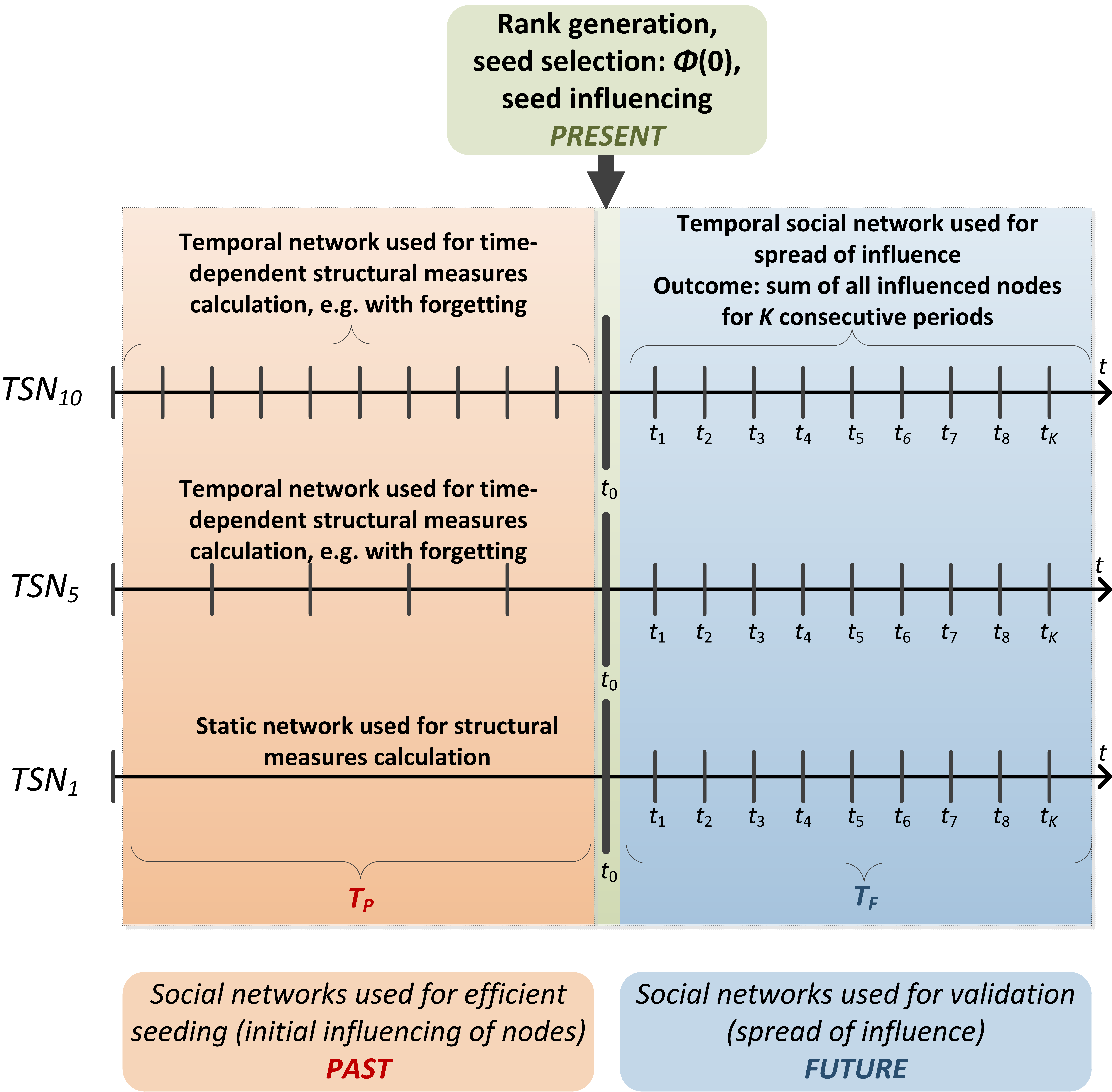}
 \caption{Seeding is performed at present based on the knowledge about past dynamics of the social network (in time $T_P$). The seed - set $\Phi(0)$ of initially influenced nodes is used to spread of influence  in the dynamic social network in the future (in time $T_F$). Three kinds of 'learning' social networks used in the experiments on seed selection are depicted one below another: $ TSN_{10} $ with 10 time windows, $ TSN_{5} $ with 5 time frames, $ TSN_{1} $ - aggregated-static (one time window).}
 \label{fig:threekinds}
\end{figure}

For that reason, the main research question one may ask is whether this dynamics existing during the spread, may be also somehow included in the seeding process. The authors intuition is that it can be done by means of usage of dynamics observed for the same community but in the previous periods - in the past. In other words, if we have some knowledge about changes in the network in the past period $T_P$, we would like to create a better seed that would enable us to influence more nodes in the dynamic network in the future $T_F$, see also Fig.~\ref{fig:threekinds}. The general problem considered in that context is what kind of networks should we use to perform better in seeding and finally in the spread outcome. Two main network kinds have been further studied: static one that aggregates equally all knowledge from the past ($TSN_1$ in Fig.~\ref{fig:threekinds}) and temporal one that splits the past period into more or less time intervals: $TSN_{10}$ with 10 equal time windows and $TSN_5$ with 5 time frames. The temporal approach corresponds to dynamic context of seeding, whereas an aggregated social network reflects typical static seeding circumstances. 

Additionally, various methods for node rankings were considered. All of them were based on different structural network measures but for temporal networks diverse forgetting methods were applied to respect new knowledge more than old one, see Section \ref{Sec:taxmes}. 

Note that the temporal approach provides a unique chance to utilize dynamics of the social network observed in the past. If this dynamics (kinds and speed of changes) is to some extent similar in the future, the time-sensitive seeding may potentially deliver better results. Efficient seeding is very important in the real world, especially in marketing campaigns - the proper selection of initial customers may significantly increase the future sell.

\newpage
\subsection{Time-dependent Users Rankings based on Structural Network Measures}
\label{Sec:taxmes}

First, we introduce three simple aggregations, which allows us to order users based on structural measures (total degree, in-degree, out-degree, betweenness, closeness) respecting all periods in the temporal social network in the accumulated way. If we consider the $m_{v_i}^l$ as a value of a given structural measure $m$ (e.g in-degree) of particular node $v_i$ in the $l$th time interval ($1\leq l\leq K$), several unnormalized aggregated measures respecting temporal aspects of node's activity in all consecutive periods can be defined as follows:

\begin{itemize}
\item Maximum
\begin{equation}
Max(v_i)=\max(\bigcup^{K}_{l=1}m_{v_i}^l)
\end{equation}
\item Minimum
\begin{equation}
Min(v_i)=\min(\bigcup^{K}_{l=1}m_{v_i}^l)
\end{equation}
\item Sum
\begin{equation}
Sum(v_i)=\sum^{K}_{l=1}m_{v_i}^l
\end{equation}
\end{itemize}

The above aggregations, however, do not make use of sequential nature of time and general phenomena that recent social relationships are likely to be more influential than old ones. Hence, the authors have introduced nine new aggregations that take into account also the "forgetting" aspect of time i.e. the value of a given structural measure in the most recent time window is the most important, while the measures value in the oldest period is the least valuable. The purpose of this, was not only to capture the dynamics of user behaviour but also to emphasize users latest activities. These new aggregations are defined in the following way:
\newpage
\begin{itemize}

\item Maximum Logarithm 
\begin{equation}
MaxLog(v_i)=\max(\bigcup^{K}_{l=1}log_{K-l+1} m_{v_i}^l)
\end{equation}
\item Minimum Logarithm
\begin{equation}
MinLog(v_i)=\min(\bigcup^{K}_{l=1}log_{K-l+1} m_{v_i}^l)
\end{equation}
\item Sum of Logarithms 
\begin{equation}
\label{Eq:SumLog}
SumLog(v_i)=\sum^{K}_{l=1}log_{K-l+1} m_{v_i}^l
\end{equation}

\item Maximum Power 
\begin{equation}
MaxPow(v_i)=\max(\bigcup^{K}_{l=1}({m_{v_i}^l})^l)
\end{equation}
\item Minimum Power
\begin{equation}
MinPow(v_i)=\min(\bigcup^{K}_{l=1}({m_{v_i}^l})^l)
\end{equation}
\item Sum of Powers 
\begin{equation}
\label{Eq:SumPower}
SumPow(v_i)=\sum^{K}_{l=1}({m_{v_i}^l})^l
\end{equation}

\item Linear Forgetting
\begin{equation}
LF(v_i)=\sum^{K}_{l=1}l m_{v_i}^l
\end{equation}
\item Hyperbolic Forgetting
\begin{equation}
\label{Eq:HypForgetting}
HF(v_i)=\sum^{K}_{l=1}\frac{1}{K-l+1}m_{v_i}^l
\end{equation}
\item Exponential Forgetting
\begin{equation}
\label{eq:ExpForgetting}
EF(v_i)=\sum^{K}_{l=1}\frac{1}{\exp(l)}m_{v_i}^l
\end{equation}
\end{itemize}

All the aggregations combined with all typical node structural measures (in-degree, out-degree, total degree, betweenness and closeness) where used to create node rankings and select the seed set for spreading the influence. However, only few of them really provided reasonable and distinct results. Finally, only six best and most representative combinations of measures and their temporal aggregation were analysed in-depth: 

\begin{itemize}
\item in-degree (InExp) and out-degree (OutExp) with exponential forgetting, Eq. \ref{eq:ExpForgetting},
\item total degree with logarithmic forgetting (TotLog), Eq. \ref{Eq:SumLog},
\item betweenness with hyperbolic forgetting (BetHyp), Eq. \ref{Eq:HypForgetting},
\item closeness with power forgetting (CloPow), Eq. \ref{Eq:SumPower},
\end{itemize}

See section \ref{seedSelection} and \ref{Sec:TemporalvsAggregated} for additional details.

In other words, nodes in the temporal social network from the past were ranked according to the time-aggregated values of their structural measures and this aggregation was performed for all component networks used for seeding, see the left part of Fig.~\ref{fig:threekinds}. Next, some top ranked nodes were used for seeding, see the middle part of Fig.~\ref{fig:threekinds}. It means that these top nodes form the initial set $\Phi(0)$ of already influenced nodes that may influence others in the following periods, see the right part of Fig.~\ref{fig:threekinds}.

\section{Experimental Setup}

\subsection{Datasets Description}
The experiments were conducted using five real-world social networks representing the communication between company employees or social services users (Table \ref{tab:datasets}). All of them were extracted from communication datasets downloaded from the Koblenz Network Collection (KONECT)\footnote{http://konect.uni-koblenz.de} repository. Each social network has timestamped edges, so it allowed to perform temporal analysis. The properties of the datasets are presented in Table \ref{tab:datasets}.

\begin{table*}[ht]
\renewcommand{\arraystretch}{1.3}
\caption{Descriptions and basic properties of used datasets}
\label{tab:datasets}
\centering
\begin{tabular}{|p{1.2cm}||p{4.2cm}||p{0.9cm}||p{1.3cm}||p{2.1cm}|}
\hline
\bfseries Dataset ID & \bfseries Network description & \bfseries \centering No. of nodes & \bfseries \centering No. of timestamped edges & \bfseries Period of communication\\
\hline
1 & E-mail communication between employees of manufacturing company\cite{michalski2011} & \centering 167 & \centering 82,927 & 2010-01-02 ... 2010-09-30\\
\hline
2 & The Enron email network\cite{klimt2004enron} & \centering 87,101 & \centering 1,147,126 & 1998-11-02 ... 2002-07-12\\
\hline
3 & Messages sent between the users of an online community of students from the University of California, Irvine\cite{konect:opsahl09} & \centering 1,899 & \centering 59,835 & 2004-04-15 ... 2004-10-26\\
\hline
4 & Facebook user to user wall posts\cite{b480} & \centering 46,952 & \centering 876,993 & 2004-09-14 ... 2009-01-22\\
\hline
5 & The reply network of the social news website Digg\cite{b565} & \centering 30,398 & \centering 87,627 & 2008-10-28 ... 2008-11-13\\
\hline
\end{tabular}
\end{table*}

\newpage
\subsection{Temporal Network Processing}
The goal of experiments was to show that the network dynamics impacts on the spread of influence process. By using the above mentioned datasets, authors created three kinds of temporal networks. Each of the communication datasets was split in two parts of equal time. The first part was three times independently processed to create three types of social networks:

\begin{itemize}
	\item a temporal social network with ten time windows of equal duration (network type $ TSN_{10} $),
	\item a temporal social network with five windows of equal duration (network type $ TSN_{5} $),
	\item a single aggregated, static network (network type $ TSN_{1} $).
\end{itemize}

In each case, the second part of the dataset was split into ten windows of equal duration, to reflect the dynamic behaviour of the network. Figure~\ref{fig:threekinds} presents how the particular network types were generated showing the learning and evaluation part of the datasets. The effect of all splitting method for the first part of the dataset on the number of nodes, edges and average in-degree in particular windows for the Facebook dataset is shown in Table \ref{tab:tsns}.

\begin{table}
\renewcommand{\arraystretch}{1.3}
\caption{Properties of social networks extracted from the Enron dataset}
\label{tab:tsns}
\centering
\begin{tabular}{|p{1.5cm}||p{1.5cm}||p{1,5cm}||p{1,5cm}||p{1.5cm}|}
\hline
\bfseries \centering  Network type & \bfseries \centering Window No. & \bfseries \centering Number of nodes & \bfseries \centering Number of edges & \bfseries Average node in-degree\\
\hline
\centering $ TSN_{10} $ & \centering 1 & \centering 63 & \centering 69 & 1.095\\ \hline
\centering $ TSN_{10} $ & \centering 2 & \centering 168 & \centering 241 & 1.435\\ \hline
\centering $ TSN_{10} $ & \centering 3 & \centering 233 & \centering 310 & 1.33\\ \hline
\centering $ TSN_{10} $ & \centering 4 & \centering 742 & \centering 1,345 & 1.813\\ \hline
\centering $ TSN_{10} $ & \centering 5 & \centering 1,057 & \centering 1,965 & 1.859\\ \hline
\centering $ TSN_{10} $ & \centering 6 & \centering 1,870 & \centering 2,926 & 1.565\\ \hline
\centering $ TSN_{10} $ & \centering 7 & \centering 4,374 & \centering 8,071 & 1.845\\ \hline
\centering $ TSN_{10} $ & \centering 8 & \centering 5,401 & \centering 11,717 & 2.169\\ \hline
\centering $ TSN_{10} $ & \centering 9 & \centering 7,708 & \centering 20,651 & 2.655\\ \hline
\centering $ TSN_{10} $ & \centering 10 & \centering 8,477 & \centering 24,361 & 2.874\\ \hline \hline
\centering $ TSN_{5} $ & \centering 1 & \centering 195 & \centering 284 & 1.456\\ \hline
\centering $ TSN_{5} $ & \centering 2 & \centering 824 & \centering 1,507 & 1.829\\ \hline
\centering $ TSN_{5} $ & \centering 3 & \centering 2,279 & \centering 4,341 & 1.905\\ \hline
\centering $ TSN_{5} $ & \centering 4 & \centering 7,625 & \centering 17,779 & 2.332\\ \hline
\centering $ TSN_{5} $ & \centering 5 & \centering 12,361 & \centering 39,143 & 3.167\\ \hline \hline
\centering $ TSN_{1} $ & \centering 1 & \centering 16,722 & \centering 55,495 & 3.319\\ \hline
\end{tabular}
\end{table}

\subsection{Influence Model Parameters}
For the linear threshold model (LT), three threshold levels $\phi_i$ assigned uniformly for all nodes $v_i$ were used: 0.33, 0.50, 0.75. It means that node $v_i$ becomes influenced if one third, a half or three fourth of its neighbours, i.e. other nodes with edges towards $v_i$, are already influenced, respectively. Naturally, this assumption is a simplification of the real-world processes, but the authors decided to keep it this way to focus on the temporal properties of networks rather than on the influence of varying threshold levels on the final outcomes. Due to the fact that the number of influenced nodes for threshold levels $\phi_i=0.33$ and $\phi_i=0.50$ was too similar among all measures and network types utilized (there was no statistically significant difference that would enable to distinguish the quality of these two approaches), the results are presented only for the highest threshold level $\phi_i=0.75$. It was the most difficult to succeed for the process of the spread of influence and it varied the results the most. To give an example, for the threshold level $\phi_i=0.33$ for all the temporal network types and for 4 out of 5 datasets, the final results of the spread of influence were the same - all possible nodes were influenced. The level $\phi_i=0.50$ introduces some differences among results but still they were not statistically significant, hence it was decided to focus only on the most challenging threshold level: $\phi_i=0.75$.

\subsection{Seed Selection}
\label{seedSelection}
In all cases, five main groups of measures were used. They were based on: (1) in-degree with exponential forgetting (InExp), (2) out-degree with exponential forgetting (OutExp), (3) total degree with logarithmic forgetting (TotLog), (4) betweenness with hyperbolic forgetting (BetHyp) and (5) closeness with power forgetting (CloPow), see also Section \ref{Sec:taxmes}. Each of these time-dependent measures was applied to different variations of social networks, i.e. static - $TSN_{1}$ and temporal - $TSN_{5}$, $TSN_{10}$. In each case, ranks for the various measures were independently generated and based on these ranks, 5\% of nodes with the highest rank value were used as seeds in the further influencing process.

\newpage
\subsection{Computations}
Computations for this experiment were conducted in Wroc{\l}aw Centre for Networking and Supercomputing\footnote{http://www.wcss.pl/en/} using R programming language\cite{R} and igraph library\cite{igraph}.

\section{Results and Discussion}

The experiments were conducted using five real-world datasets. For all of them results revealed that the influence of network type used for choosing initial seeds is significant for the total number of influenced nodes. In particular, static context of seeding (network $TSN_{1}$) was confronted against temporal approach (networks: $TSN_{5}$ and $TSN_{10}$). Additionally, various time-dependent structural measures were analysed and finally some of them were tested in all three contexts.

\subsection{Time-dependent Structural Measures}
\label{Sec:ResultMeasures}

First, it was evaluated by means of statistical tests (Friedman\cite{friedmantest} and Nemenyi\cite{nemenyi}) which variations in each measure group perform best. Results showed that in all cases variations using some temporal aspects were performing the best and finally the following measures were analysed more in-depth, see Tab.~\ref{tab:rawnumbers}, \ref{tab:friedman} and \ref{tab:nemenyi}. It referred the following measures:

\begin{itemize}
	\item exponential forgetting for out-degree (OutExp) and in-degree (InExp),
	\item logarithmic forgetting for total degree (TotLog),
	\item hyperbolic forgetting for betweenness (BetHyp),
	\item power forgetting for closeness (CloPow).
\end{itemize}

\begin{figure}
\includegraphics[width=\columnwidth]{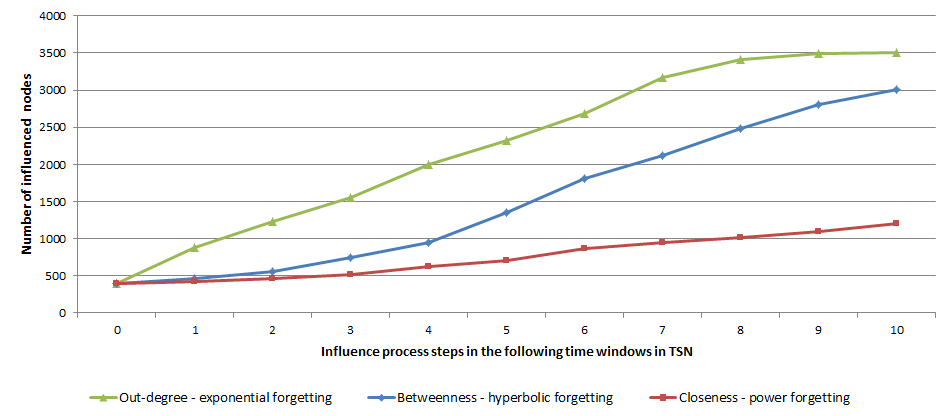}
 \caption{The number of influenced nodes for the fourth dataset (Facebook), network type $ TSN_{10} $ and the threshold level $\phi_i=0.75$ for best performing measures of each base type (total degree, betweenness, closeness).}
 \label{fig:numofinf}
\end{figure}

\begin{figure}
        \begin{subfigure}[b]{0.5\textwidth}
                \includegraphics[width=\textwidth]{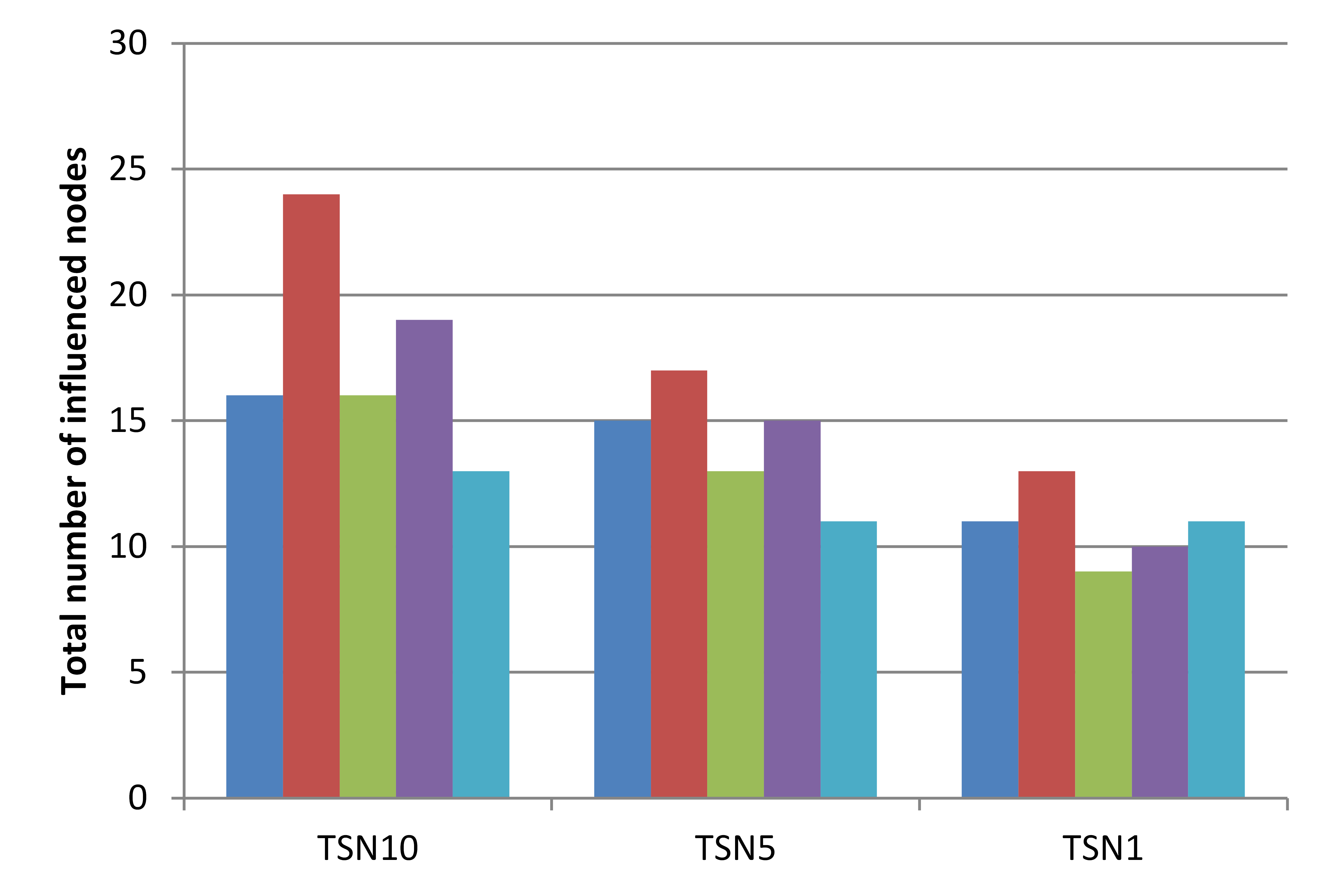}
                \caption{\small \sl Manufacturing company}
                \label{fig:allresults_manuf}
        \end{subfigure}
        \begin{subfigure}[b]{0.5\textwidth}
                \includegraphics[width=\textwidth]{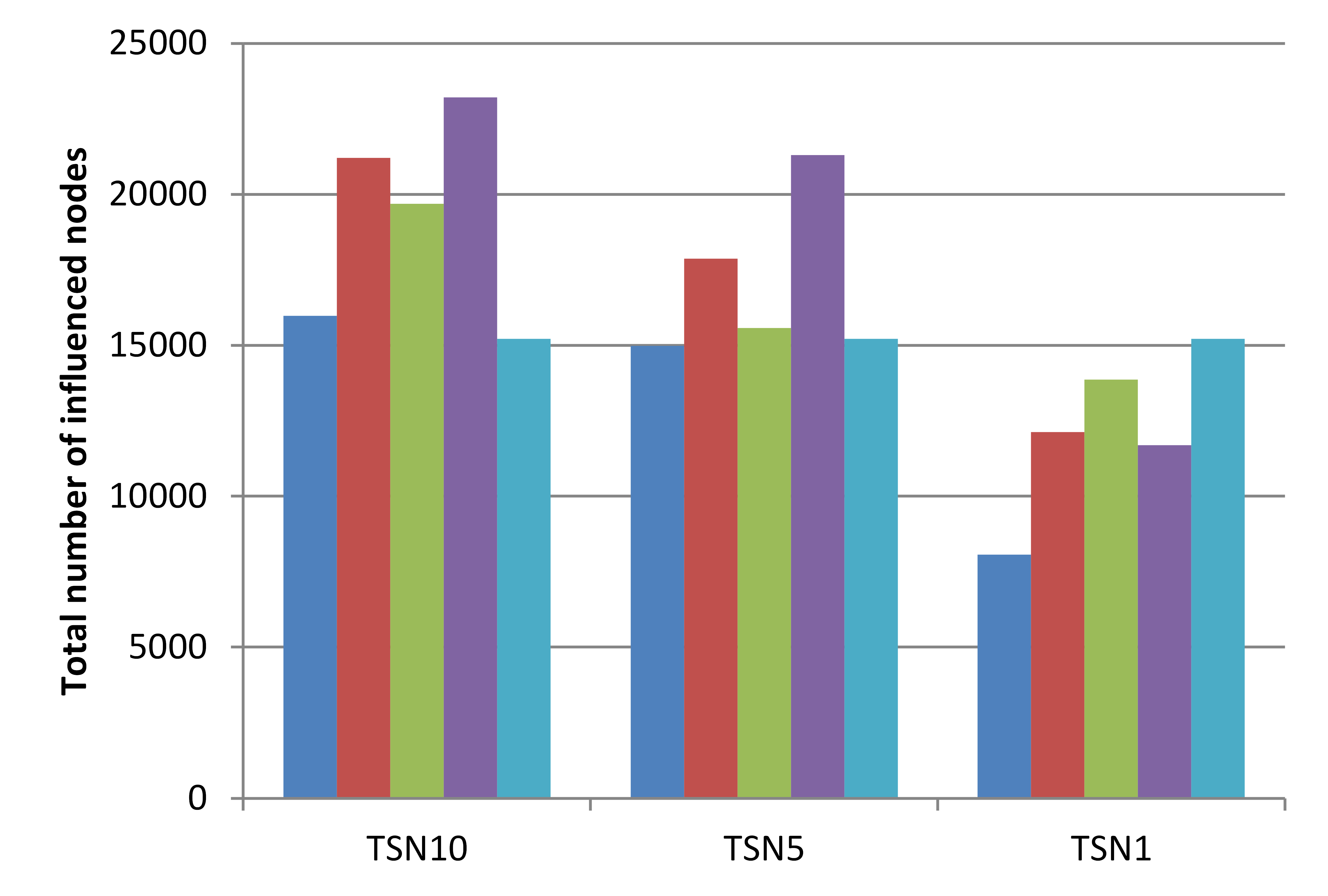}
                \caption{\small \sl Enron}
                \label{fig:allresults_enron}
        \end{subfigure}
        \begin{subfigure}[b]{0.5\textwidth}
                \includegraphics[width=\textwidth]{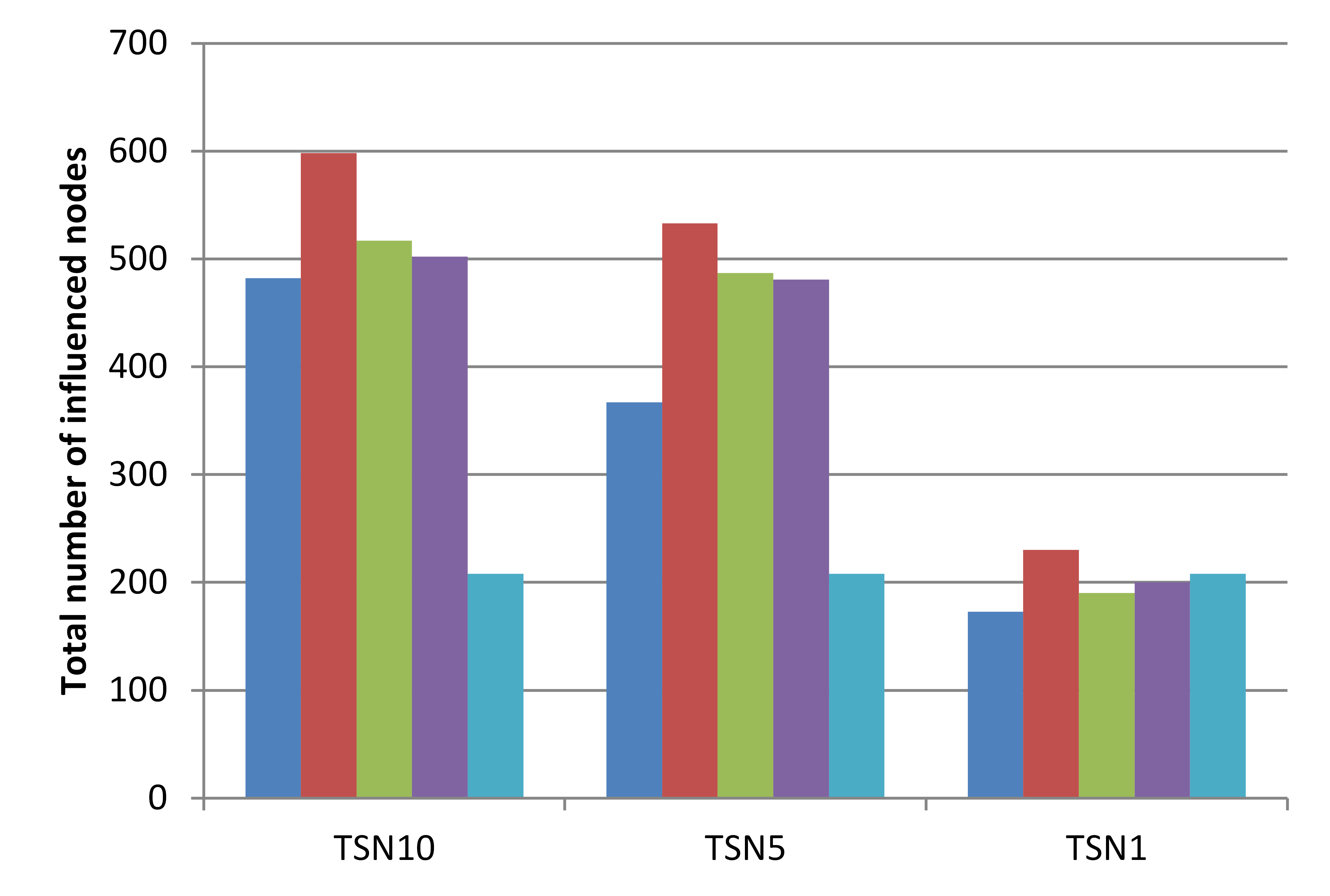}
                \caption{\small \sl University of California}
                \label{fig:allresults_uoc}
        \end{subfigure}
        \begin{subfigure}[b]{0.5\textwidth}
                \includegraphics[width=\textwidth]{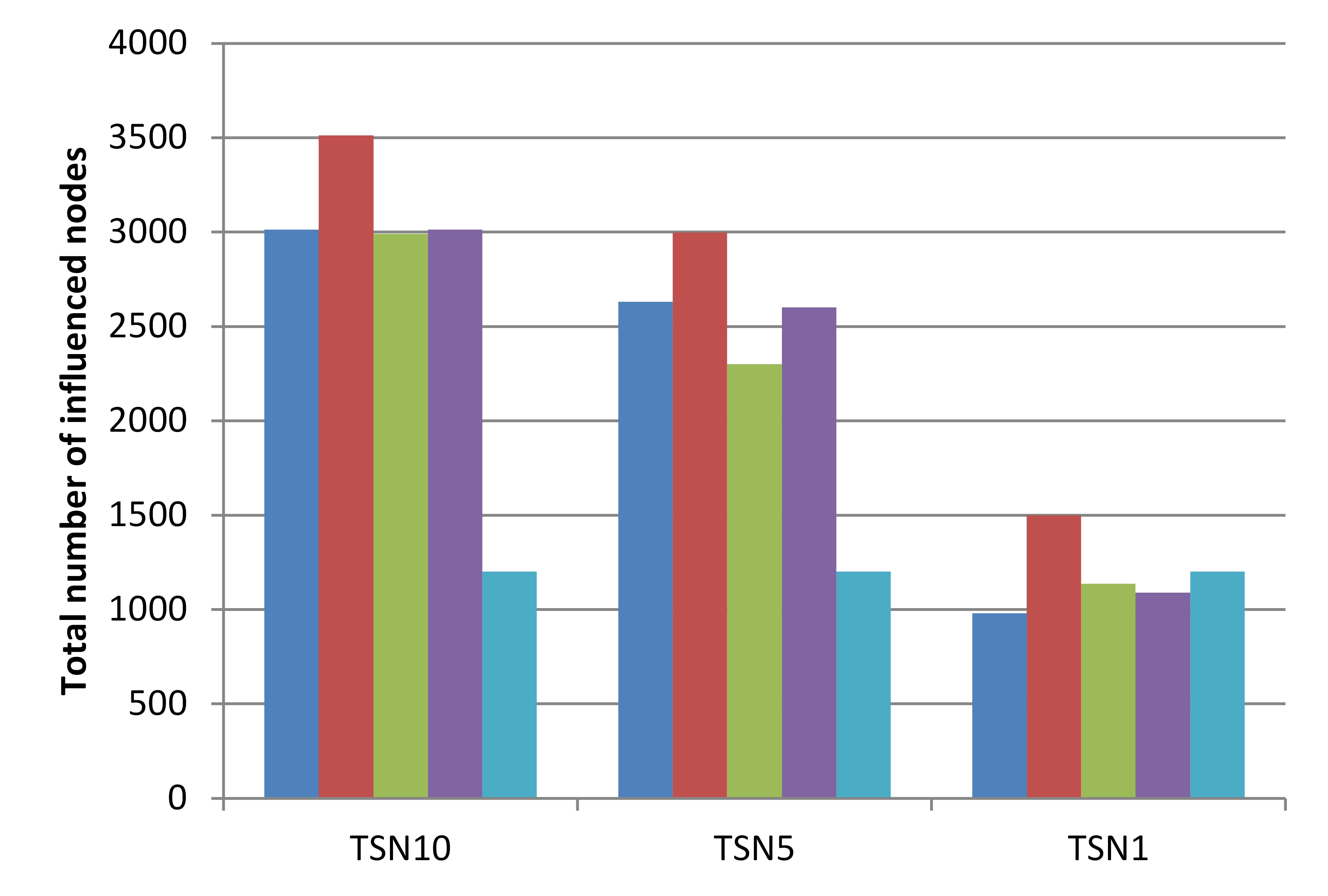}
                \caption{\small \sl Facebook}
                \label{fig:allresults_facebook}
        \end{subfigure}
        \begin{subfigure}[b]{1.0\textwidth}
		        \centering
                \includegraphics[width=0.5\textwidth]{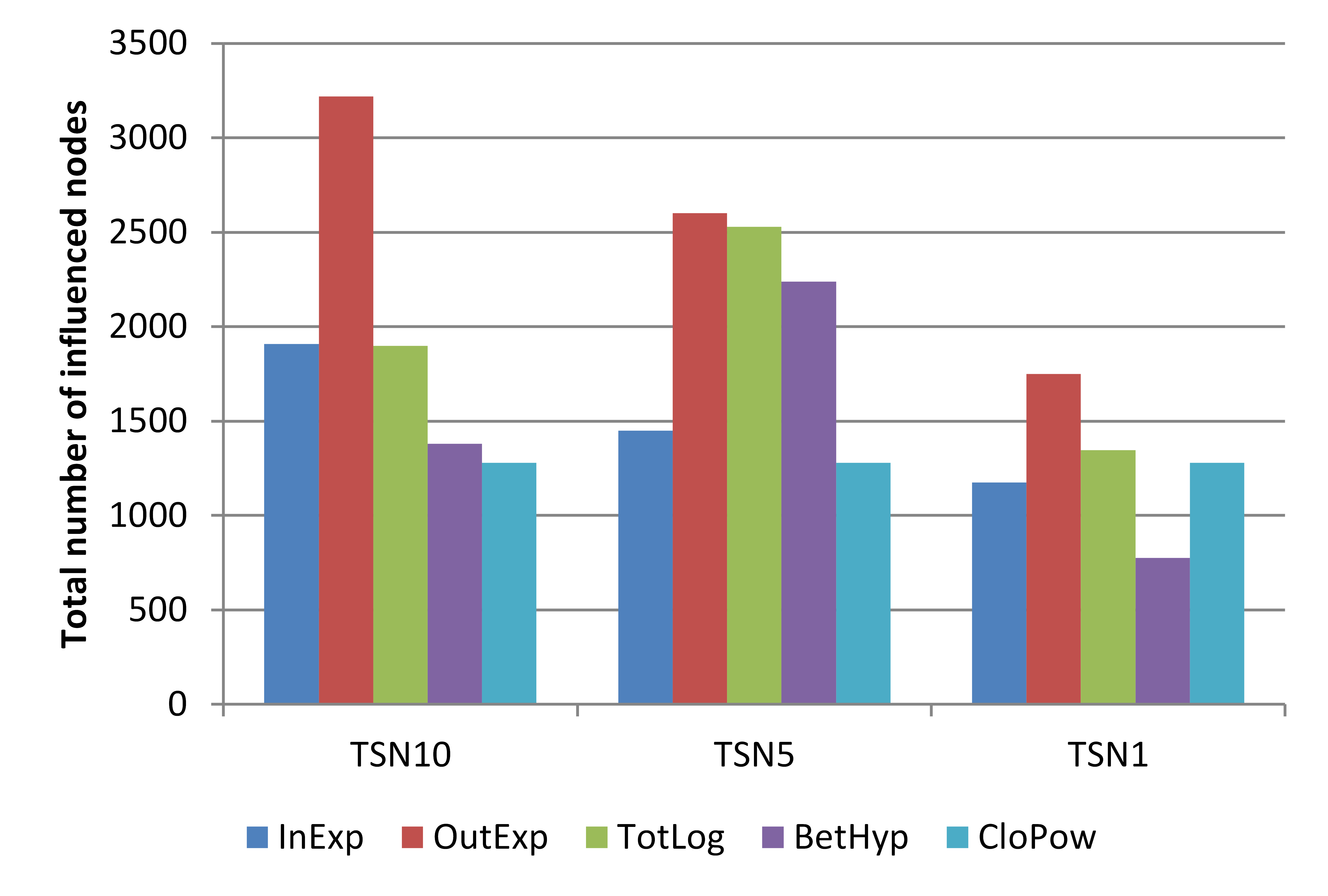}
                \caption{\small \sl Digg}        
                \label{fig:allresults_digg}
        \end{subfigure}
        \caption{The total number of influenced nodes for all networks and structural measures used for seeding as well as for different datasets,  the threshold level $\phi_i=0.75 $}
        \label{fig:allresults}
\end{figure}

The results for all the datasets and best measures are presented in Fig.~\ref{fig:allresults}. To present how the process followed in time, the results for the Facebook dataset are depicted in Fig.~\ref{fig:numofinf} with more precise data for the same dataset presented in Tab.~\ref{tab:rawnumbers}.

\begin{table}
  \centering
  \caption{Total number of the influenced nodes for different seeding strategies on different network types; results are presented for the Facebook dataset and the threshold level $\phi_i=0.75$}
    \begin{tabular}{|p{3cm}|c|c|c|}
    \hline
    Measure type & $ TSN_{10} $ & $ TSN_{5} $ & $ TSN_{1} $ \\
    \hline
    InExp & 3,012 & 2,630 & 980 \\
    OutExp & \textbf{3,512} & \textbf{2,998} & \textbf{1,500} \\
	TotLog & 2,991 & 2,300 & 1,136 \\
    BetHyp & 3,014 & 2,600 & 1,090 \\
    CloPow & 1,200 & 1,200 & 1,200 \\
    Random & 1,682 & 1,151 & 940 \\
    Random$_{\textrm{freq}}$ & 2,241 & 1,923 & 1,132 \\
    \hline
    \end{tabular}
  \label{tab:rawnumbers}
\end{table}

All the above basic heuristics performed better than two random algorithms of ranking. After analysing all the results, the best performing group of measures for most of the datasets was the one based on out-degree - in just one case betweenness was performing better. It referred almost all datasets and temporal contexts (number of splits). In only one dataset - Enron (Fig.~\ref{fig:allresults}b), the betweeness measure with hyperbolic forgetting was the best but out-degree occupied the second position.

\subsection{Temporal vs. Aggregated Networks}
\label{Sec:TemporalvsAggregated}

Next, these variants were tested on three different networks: temporal ($ TSN_{10} $, $ TSN_{5} $) and aggregated ($ TSN_{1} $) to see whether there are any statistically significant differences between the total number of influenced nodes at the end of the process. As it was mentioned before, separate experimental temporal networks were built for five real world datasets. As a baseline, two random seeding strategies were utilized: (1) an algorithm, which draws nodes with the same probability ($ Random $) and (2) another one that selects nodes based on their frequency of occurrence in particular time windows ($ Random_{freq} $), i.e. the node that occurs more frequently in all time windows before the seed selection will have a greater chance to be chosen as a seed. Both random algorithms were run a hundred times and the results were averaged. Table \ref{tab:rawnumbers} shows how particular algorithms perform on both temporal and static networks for the Facebook dataset.

\begin{table}
  \centering
  \caption{The average percentage of neighbours exchanged by initially chosen seeds in each time window during the influence process - comparison of seeds chosen using OutExp measure and random ones. Network of University of California, $ TSN_{10} $.}
    \begin{tabular}{|c|c|c|c|}
    \hline
    Time window & OutExp & Random & Std. deviation (random) \\
    \hline
    $ t_1 $ & 44.04\% & 10.80\% & 3.14\% \\
    $ t_2 $ & 30.79\% & 8.63\% & 2.70\% \\
	$ t_3 $ & 28.99\% & 9.29\% & 2.57\% \\
    $ t_4 $ & 29.44\% & 10.00\% & 3.03\% \\
    $ t_5 $ & 24.22\% & 6.69\% & 2.22\% \\
    $ t_6 $ & 23.56\% & 8.09\% & 2.27\% \\
    $ t_7 $ & 19.77\% & 6.38\% & 2.21\% \\
    $ t_8 $ & 21.59\% & 5.16\% & 1.93\% \\
    $ t_9 $ & 14.04\% & 4.35\% & 1.80\% \\
    $ t_{10} $ & 17.97\% & 6.16\% & 2.31\% \\
    \hline
    \hline
    \textbf{Average} & \textbf{25.44}\% & \textbf{7.56}\% & \textbf{2.14}\% \\
    \hline
    \end{tabular}
  \label{tab:neighchange}
\end{table}

Authors also performed the analysis how the neighbourhood of chosen seeds changes in time during the influence process. In Table \ref{tab:neighchange} it is presented the comparison of the best performing method (OutExp) and random seeds during the influence stage for University of California datatset and $ TSN_{10} $. For instance, seeds chosen by OutExp method changed their neighbourhood for 44.04\% after the first time window, whereas random seeds had only 10.80\% change in their neighbourhood. Here, random method seed selection was repeated 100 times and results were averaged.

To confirm the statistical significance of results, typical statistical tests were applied, based on NxM Friedman\cite{friedmantest} test and multiple post-hoc comparisons. The average ranks as well as the Friedman p-values showing how particular algorithms performed among all test datasets by using different aggregations are presented in Table \ref{tab:friedman}. It is clearly visible that the ranking of network granularities is always the same for all strategies (measures) except closeness (CloPow): the best is the temporal network with the highest granularity - with 10 time windows ($TSN_{10}$), the second is the one with less periods ($TSN_{5}$) and the worst, third is the static aggregated network $TSN_{1}$. These results are statistically significant (p-value $<$ 5\%) among all datasets and lead to the crucial conclusion that the greater granularity (the greater the number of consecutive periods), the better. This is so because, we are able to extract more information from raw dataset instead aggregating all of it into one link between two nodes in static network like $TSN_{1}$, and in consequence to model network dynamics and user behaviour much closer to the actual, true network dynamics.

\begin{table}
  \centering
  \caption{Ranks of seed strategies based on various measures using Friedman test for different network types, all datasets combined, threshold $\phi_i=0.75$}
    \begin{tabular}{|p{3cm}|c|c|c|c|c|}
    \hline
    Network type & InExp & OutExp & TotLog & BetHyp & CloPow \\
    \hline
    $ TSN_{10} $ & \textbf{1.1} & \textbf{1} & \textbf{1.2} & \textbf{1} & 1.8 \\
    $ TSN_{5} $ & 1.9 & 2 & 1.8 & 2 & 2 \\
	$ TSN_{1} $ & 3 & 3 & 3 & 3 & 2.2 \\
	\hline
    Friedman p-value & 0.01 & 0.006 & 0.01 & 0.007 & 0.81 \\
    \hline
    \end{tabular}
  \label{tab:friedman}
\end{table}

An additional post-hoc analysis was also done by using Nemenyi\cite{nemenyi} pairwise procedure and the results are presented in Table \ref{tab:nemenyi}. This analysis showed that the most significant difference is seen for the comparison of  $TSN_{10}$ and  $TSN_{1}$ - the best with the worst. It also confirms the general conclusion that the greater difference in granularity, the greater gain in outcome (much more nodes are influenced). Same conclusions were drawn by using different post-hoc procedures, like Shaffer\cite{shaffer}, Bergman\cite{bergmann} and Holm\cite{holm}. The only measure that was not revealing the same phenomenon was closeness, but after further analysis authors found out that nodes had very similar values of measures based on closeness in every network type, so in terms of the spread of influence process outcomes no differences were seen - almost always the same seeds were chosen.

\begin{table}
  \centering
  \caption{Adjusted p-values of the post-hoc Nemenyi procedure for different network types, all  datasets combined, threshold $\phi_i=0.75$}
    \begin{tabular}{|p{3.2cm}|c|c|c|c|c|}
    \hline
    Network type & InExp & OutExp & TotLog & BetHyp & CloPow \\
    \hline
    $ TSN_{10} $ vs. $ TSN_{1} $ & \bfseries 0.008 & \bfseries 0.005 & \bfseries 0.013 & \bfseries 0.002 & 1.581268 \\
    $ TSN_{5} $ vs. $ TSN_{1}  $ & 0.246 & 0.341539 & 0.173339 & 0.113846 & 2.255489 \\
	$ TSN_{10} $ vs. $ TSN_{5}  $ & 0.618 & 0.341539 & 1.028345 & 0.113846 & 2.255489 \\
    \hline
    \end{tabular}
  \label{tab:nemenyi}
\end{table}

\subsection{Discussion}
\label{Sec:Discussion}
Results reveal that indeed for the aggregated (static) network, i.e. $TSN_{1}$, the total number of the influenced nodes is the lowest (the right group of bars in Fig.~\ref{fig:allresults}) and the best performing network type is the one with the biggest number of time windows, i.e. $TSN_{10}$  - the left hand side group of bars, Fig.~\ref{fig:allresults}. Overall, the final number of influenced nodes for the 10-windows networks ($TSN_{10}$) was about double as much as for a single network $TSN_{1}$, see Fig.~\ref{fig:allresults}. It confirms our initial hypothesis that using dynamic network we are able to better utilize the information in original data and finally select better seeds.

What is more, the greater granularity, the better chance to choose the proper seeds, especially if taking time into consideration by means of time-dependent measures, such as based on linear forgetting. When trying to explain this phenomenon, once again the intuition is suggesting that the increasing granularity is helpful in terms of better representation of the network dynamics, so the sensitivity of the introduced measures increases -- they reflect dynamics to a greater extent. As studies\cite{qsd} show, the network dynamics is very prone to the size of the time window, so in this case the more detailed representation of facts the better, because very short and intensive actions, such as bursts, will be captured and represented better, without being averaged by longer time periods.

 \begin{figure}
 \centering
 \includegraphics[scale=.4]{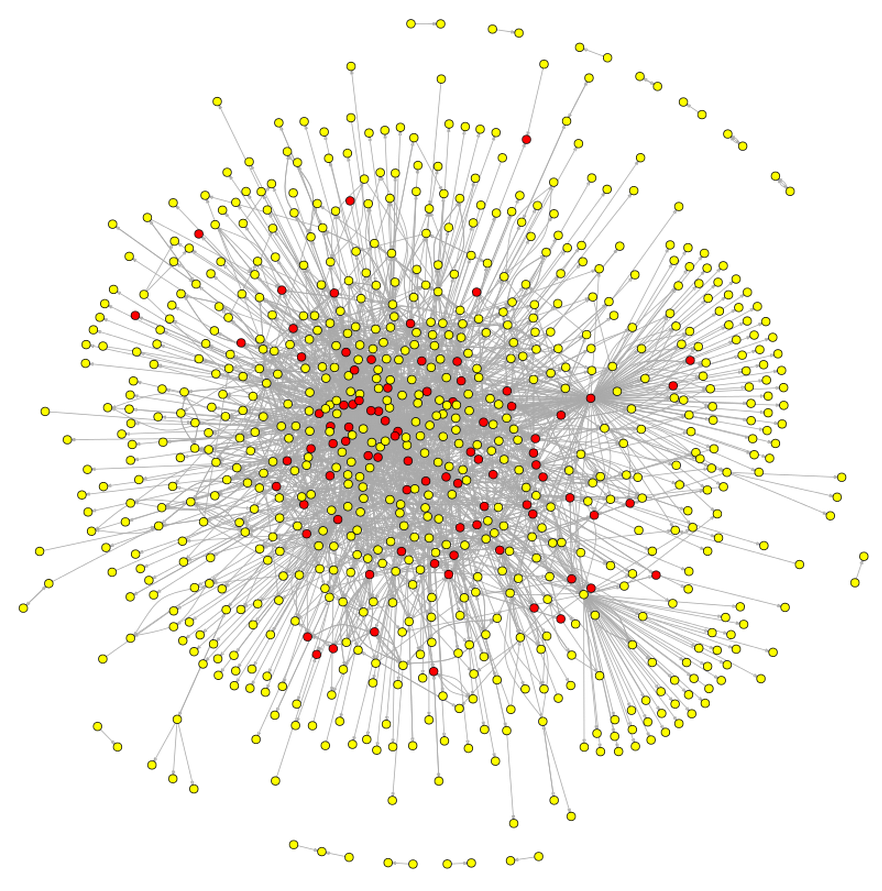}
 \caption{The position of seeds (marked red) chosen by using OutExp measure for the University of California dataset and $ TSN_{10} $. For the presentation purpose links for ten time windows of the influence (evaluation) process were aggregated to show the complete link structure, not just the single window.}
 \label{fig:network}
\end{figure}

Coming back to the results, we have noticed that out-degree-based measures (OutExp) are performing better than others. To give more in-depth explanation, it is necessary to understand the basics of the LT model. It is worth to remember, that in directed social networks a person becomes influenced if a fraction of neighbours contacting with them is already influenced. It results with the situation that a node with the high out-degree becomes an influencer for the high number of nodes; these nodes have most probably a small degree. On the other hand, if we look at the betweenness (BetHyp), its high value means that there are a lot of shortest paths going through that node, so it could be a bridge connecting two or more parts of the network. However, this might not be enough when we are using the LT model, since the node may have only few neighbours. Of course, such a node could be essential, if one part of the network tends to influence another one and that is why the betweenness-based measures also achieved quite good results in experiments. The question for discussion and future work is whether those two measures (out-degree and betweenness) should not be somehow combined to create even better ranking for seed selection.

Taking into account all time windows, the average exchange of neighbours of the chosen seeds was more than three times higher than for the random seeds, see Table \ref{tab:neighchange}. It leads to the conclusion that the high values of the OutExp measure over all time frames was obtained not by constant intense contacts with the same group of nodes over long time, but by exchanging the dense neighbourhoods from window to window as well -- they quite frequently swapped the neighbours. As a result, during the influence part, these nodes were able to influence new neighbours in consecutive time windows giving better final results. The in-depth analysis of the position of seeds also revealed that in all datasets, these nodes were located not on boundaries of the network, but close to its center. Moreover, these seeds were rarely located close to each other, see Figure~\ref{fig:network}. 

\section{Conclusions}
Selection of nodes that are initially influenced and next 'are used' to influence the others in the social community (seeding) is one of the essential problems in analysis of spread of influence. In the real world, social networks continuously evolve and change their node and connection set contents. A proper choice of the most promising nodes becomes extremely difficult in such dynamic environment. 

The main goal of this paper was to confront static context of seeding against dynamic one. In the static approach, the social network used for seeding aggregates all data from the past in the form of the single social network. A new idea introduced in this paper -- a temporal context -- enables, in turn, to respect network dynamics observed in the past and make use of it for seeding and spread of influence within the temporal social network in the future. This unique concept is based on calculation of structural measures for a series of historical networks, having also in mind the sequential nature of time. As a result, the most recent network snapshots influence the final node ranking more than the older ones. This time-dependent ranking in a sense encapsulates the past dynamics of the social network.

The experimental studies performed on five real data sets on human communication facilitated creation of both static and temporal social networks used for seeding. The raw results as well as statistical tests have revealed the crucial paper finding: better results (more influenced people) may be achieved if seeding is carried out on known temporal social networks rather than in the static environment. Moreover, the higher granularity of periods in the network, the greater outcome and it has been observed for almost all cases. The final number of influenced nodes for the 10-windows networks was about double as much as for the single network.

Additionally, various time-dependent structural measures were also analysed. The experiments have shown that out-degree node measure with exponential forgetting performed best for most contexts. In only one dataset - Enron, the betweenness measure with hyperbolic forgetting was the best.

All the above new findings are quite coherent for all data sets analysed, which represent different public available social networks, see Table \ref{tab:datasets} and have been confirmed by statistical tests. However, we cannot state that the general rules discovered in the paper will be valid for all data sets and all environments. We also encourage the other independent researchers and also practitioners to validate them for other measures, other data sets, other settings and other than linear threshold models for spread of influence. 

\begin{acknowledgment}
This work was partially supported by the fellowship co-financed by the European Union within the European Social Fund, by the European Commission under the 7th Framework Programme, Coordination and Support Action, Grant Agreement Number 316097, ENGINE - European research centre of Network intelliGence for INnovation Enhancement (http://engine.pwr.wroc.pl/) and by The National Science Centre, the decision no. DEC-2013/09/B/ST6/02317. The calculations were carried out in Wroclaw Centre for Networking and Supercomputing (http://www.wcss.wroc.pl), grant No 177.
\end{acknowledgment}


\newpage
\begin{thebibliography}{99}
 \bibitem{bergmann} Bergmann, B., and Gerhard H.,
 \newblock "Improvements of general multiple test procedures for redundant systems of hypotheses.",
 \newblock {\em Multiple Hypothesenprüfung/Multiple Hypotheses Testing}, Springer Berlin Heidelberg, pp. 100-115, 1988.
 
 \bibitem{chen2010scalable} Chen, W. and Yuan, Y. and Zhang, L.,
 \newblock "Scalable influence maximization in social networks under the linear threshold model",
 \newblock in {\em Proceedings of 2010 IEEE 10th International Conference on Data Mining},
 IEEE Computer Society, pp. 88-97, 2010.

 \bibitem{b565} Choudhury, M. and Sundaram, H. and John, A. and Seligmann, D.D.,
 \newblock "Social Synchrony: Predicting Mimicry of User Actions in Online Social Media",
 \newblock in {\em Proc. Int. Conf. on Computational Science and Engineering},
 pp. 151-158, 2009.
  
 \bibitem{clifford1973votermodel} Cliffor, P. and Sudbury, A.,
 \newblock "A model for spatial conflict",
 \newblock {\em Biometrika vol. 60, no. 3}, pp. 581-588, 1973.

 \bibitem{igraph} Csardi, G. and Nepusz, T.,
 \newblock "The igraph software package for complex network research",
 \newblock {\em InterJournal, vol. Complex Systems}, 2006.

 \bibitem{Even-DarShapira2007} Even-Dar, E. and Shapira, A.,
 \newblock "A note on maximizing the spread of influence in social networks",
 \newblock {\em Network, vol. 111, no. 4, ch. 27}{ ed. Deng, X. and Graham, F.}, pp. 281-286, 2007.

 \bibitem{Freeman1977} Freeman, L.C.,
 \newblock "Set of Measures of Centrality Based on Betweenness",
 \newblock {\em Sociometry, vol. 40, no. 1}, pp. 35-41, 1977.

 \bibitem{friedmantest} Friedman, M.,
 \newblock "The use of ranks to avoid the assumption of normality implicit in the analysis of variance.",
 \newblock {\em Journal of the American Statistical Association 32.200}, pp. 675-701, 1937.
 
 \bibitem{goldenberg2001talk} Goldenberg, J. and Libai, B. and Muller, E.,
 \newblock "Talk of the network: A complex systems look at the underlying process of word-of-mouth",
 \newblock {\em Marketing letters, vol. 12, no. 3}, pp. 211-223, 2001.

 \bibitem{Gomez-Rodriguez_Leskovec_Krause_2010} Gomez-Rodriguez, M. and Leskovec, J. and Krause, A.,
 \newblock "Inferring Networks of Diffusion and Influence",
 \newblock in {\em Proceedings of the 16th ACM SIGKDD international conference on Knowledge discovery and data mining KDD 10, vol. 5, no. 4},
 IEEE Computer Society, pp. 1019-1028, 2010.
 
 \bibitem{goyal2011data} Goyal, A. and Bonchi, F. and Lakshmanan, L. VS,
 \newblock "A data-based approach to social influence maximization",
 \newblock {\em In Proceedings of the VLDB Endowment}, vol. 5, no. 1, pp. 73-84, 2011.

 \bibitem{GoyalBonchiLakshmananVenkatasubramanian2012} Goyal, A. and Bonchi, F. and Lakshmanan, L.V.S. and Venkatasubramanian, S.,
 \newblock "On minimizing budget and time in influence propagation over social networks"
 \newblock {\em Social Network Analysis and Mining, vol. 3, no. 2}, pp. 179-192, 2013.
 
 \bibitem{GoyalLuLakshmanan2011} Goyal, A. and Lu, W. and Lakshmanan, L.V.S.
 \newblock "SIMPATH: An Efficient Algorithm for Influence Maximization under the Linear Threshold Model",
 \newblock in {\em Proceedings of 11th IEEE International Conference on Data Mining},
 IEEE Computer Society, pp. 211-220, 2011.
  
 \bibitem{Granovetter1978} Granovetter, M.,
 \newblock "Threshold Models of Collective Behavior",
 \newblock {\em American Journal of Sociology, vol. 83, no. 6}, pp. 1420-1443, 1978.

 \bibitem{holm} Holm, S.,
 \newblock "A simple sequentially rejective multiple test procedure.",
 \newblock {\em Scandinavian journal of statistics}, pp. 65-70, 1969.

 \bibitem{holme2012temporal} Holme, P. and Saram{\"a}ki, J.,
 \newblock "Temporal networks",
 \newblock {\em Physics reports, vol. 519, no. 3}{ ed. Deng, X. and Graham, F.}, pp. 97-125, 2012.

 \bibitem{jankowski2013} Jankowski, J. and Michalski, R. and Kazienko, P.,
 \newblock "Compensatory Seeding in Networks with Varying Availability of Nodes",
 \newblock in {\em Proc. of 2013 IEEE/ACM International Conference on Advances in Social Networks Analysis and Mining ASONAM 2013},
 pp. 1242-1249, 2013.

 \bibitem{karimi2013threshold} Karimi, F. and Holme, P.,
 \newblock "Threshold model of cascades in empirical temporal networks",
 \newblock {\em Physica A: Statistical Mechanics and its Applications, vol. 392}, pp. 3476-3483, 2013.

 \bibitem{karsai2011small} Karsai, M. and Kivel{\"a}, M. and Pan, R.K. and Kaski, K. and Kert{\'e}sz, J. and Barab{\'a}si, A-L. and Saram{\"a}ki, J.,
 \newblock "Small but slow world: How network topology and burstiness slow down spreading",
 \newblock {\em Physical Review E, vol. 83, no. 2}, 2011.
  
 \bibitem{kazienko2012} Kazienko, P. and Kajdanowicz, T.,
 \newblock "Label-dependent node classification in the network",
 \newblock {\em Neurocomputing}, vol. 75, pp. 199-209, 2012.
 
 \bibitem{KempeKleinbergTardos2003} Kempe, D. and Kleinberg, J. and Tardos, E.,
 \newblock "Maximizing the spread of influence through a social network",
 \newblock in {\em Proceedings of the ninth ACM SIGKDD international conference on Knowledge discovery and data mining - KDD  2003}, ACM Press, pp. 137-146, 2003.

 \bibitem{klimt2004enron} Klimt, B. and Yang, Y.,
 \newblock "The enron corpus: A new dataset for email classification research",
 \newblock in {\em Proceedings of ECML 2004 - European Conference on Machine learning},
 Springer, pp. 217-226, 2004.
 
 \bibitem{Gueorgi2006} Kossinets, G. and Watts, D.J.,
 \newblock "Empirical analysis of an evolving social network",
 \newblock {\em Science, vol. 311, no. 5757}{ ed. Deng, X. and Graham, F.}, pp. 88-90, 2006.

 \bibitem{krolprop} Kr{\'o}l, D.,
 \newblock "On Modelling Social Propagation Phenomenon",
 \newblock {\em Lecture Notes in Computer Science, vol. 8398}, Springer Verlag, pp. 227-236, 2014. 
  
 \bibitem{liben} Liben-Nowell, D., and Kleinberg, J.,
 \newblock "The link prediction problem for social networks."
 \newblock {\em Journal of the American society for information science and technology 58.7}, pp. 1019-1031, 2007.

 \bibitem{MasudaHolme2013} Masuda, N. and Holme, P.,
 \newblock "Predicting and controlling infectious disease epidemics using temporal networks",
 \newblock {\em F1000prime reports, vol. 5}, 2013.
  
 \bibitem{mathioudakis2011sparsification} Mathioudakis, M. and Bonchi, F. and Castillo, C. and Gionis, A. and Ukkonen, A.,
 \newblock "Sparsification of influence networks"
 \newblock in {\em Proceedings of the 17th ACM SIGKDD international conference on Knowledge discovery and data mining}, ACM Press, pp. 529-537, 2011.

 \bibitem{qsd} Michalski, R. and Br\'{o}dka, P. and Kazienko, P. and Juszczyszyn, K.,
 \newblock "Quantifying social network dynamics",
 \newblock {\em In Proceedings of the 4th Conference on Computational Aspects of Social Networks (CASoN)}, IEEE Computer Society, pp. 69-74, 2012.

 \bibitem{convince} Michalski, R. and Kazienko, P. and Jankowski, J.,
 \newblock "Convince a Dozen More and Succeed--The Influence in Multi-layered Social Networks",
 \newblock {\em In Proceedings of the International Conference on Signal-Image Technology \& Internet-Based Systems (SITIS 2013)}, IEEE Computer Society, pp. 499-505, 2013.

 \bibitem{michalski2011} Michalski, R. and Palus, S. and Kazienko, P.,
 \newblock "Matching Organizational Structure and Social Network Extracted from Email Communication",
 \newblock {\em Lecture Notes in Business Information Processing, vol. 87}, pp. 197-206, 2011.
 
 \bibitem{moon1965cliques} Moon, J.W. and Moser, L.,
 \newblock "On cliques in graphs",
 \newblock {\em Israel journal of Mathematics}, pp. 23-28, 1965.

 \bibitem{nemenyi} Nemenyi, P.,
 \newblock "Distribution-free multiple comparisons",
 \newblock {\em Dissemination at Princeton University}, 1963.

 \bibitem{konect:opsahl09} Opsahl, T. and Panzarasa, P.,
 \newblock "Clustering in Weighted Networks",
 \newblock {\em Social Networks, vol. 31, no. 2}, pp. 155-163, 2009.
 
 \bibitem{PallaBarabasiVicsek2007} Palla, G, and Barabási, A-L. and Vicsek, T.,
 \newblock "Quantifying social group evolution",
 \newblock {\em Nature}, 2007.

 \bibitem{prell2011social} Prell, C.,
 \newblock {\em Social network analysis: History, theory and methodology},
 \newblock Sage Publications Limited, 2011.

 \bibitem{R} R Development Core Team,
 \newblock {\em R: A Language and Environment for Statistical Computing},
 \newblock R Foundation for Statistical Computing, 2011. 
 
 \bibitem{rogers2010diffusion} Rogers, E.M.,
 \newblock {\em Diffusion of innovations},
 \newblock Simon and Schuster, 2010. 

 \bibitem{SaitoNakanoKimuraLovrekHowlettJain2008} Saito, K. and Nakano, R. and Kimura, M. and Lovrek, I. and Howlett, R. and Jain, L.,
 \newblock "Prediction of Information Diffusion Probabilities for Independent Cascade Model",
 \newblock in {\em KnowledgeBased Intelligent Information and Engineering Systems}{ ed. Lovrek, I. and Howlett, R.J. and Jain, L.}, Springer Verlag, pp. 67-75, 2008.

 \bibitem{shaffer} Shaffer, JP,
 \newblock "Multiple hypothesis testing."
 \newblock {\em Annual review of psychology 46.1 (1995)}, pp. 561-584.

 \bibitem{spira1975finding} Spira, P.M. and Pan, A.,
 \newblock "On finding and updating spanning trees and shortest paths",
 \newblock {\em IAM Journal on Computing, vol. 4, no. 3}, pp. 375-380, 1975.

 \bibitem{b480} Viswanath, B. and Mislove, A. and Cha, M. and Gummadi, K.P.,
 \newblock "On the Evolution of User Interaction in Facebook",
 \newblock in {\em Proc. Workshop on Online Social Networks},
 Springer, pp. 37-42, 2009.
  
\end{thebibliography}
\end{document}